\documentclass[11pt,a4paper]{article}

\usepackage[margin=1in]{geometry}
\usepackage{graphicx}
\usepackage{amsmath,amssymb}
\usepackage{xcolor}
\usepackage{listings}
\usepackage[colorlinks=true, urlcolor=blue, linkcolor=blue, citecolor=blue]{hyperref}
\usepackage{titlesec}

\titleformat{\section}{\large\bfseries}{\thesection.}{0.5em}{}
\titleformat{\subsection}{\normalsize\bfseries}{\thesubsection.}{0.5em}{}

\begin{document}
	
	\title{\textbf{Co-Authoring with AI: \\ How I Wrote a Physics Paper About AI, Using AI}}
	\author{Yi Zhou \\ \textit{Institute of Physics, Chinese Academy of Sciences, Beijing 100190, China}}
	\date{\today}
	
	\maketitle
	
	\begin{abstract}
		The rapid integration of Large Language Models (LLMs) into scientific writing fundamentally challenges traditional definitions of authorship, responsibility, and scientific integrity. As researchers transition from using computers as deterministic tools to managing them as ``virtual collaborators,'' the nature of human contribution must be re-evaluated. Using the drafting process of a recent computational physics manuscript as a case study, this essay explores the indispensable role of the Human-in-the-Loop (HITL). We demonstrate that while AI excels at structural organization and syntax generation, the human author bears the ultimate responsibility for enforcing rigorous physical logic, maintaining academic diplomacy, and anticipating peer-review critiques. In this paradigm, the human contribution shifts from writing boilerplate text to acting as a Principal Investigator who actively mentors and steers the AI's reasoning. To ensure accountability and preserve the integrity of the scientific record in this new era, I argue that the community must mandate the publication of full, unedited AI interaction transcripts as standard supplementary material.
	\end{abstract}
	
	\vspace{0.5cm}
	
	\section{The Paradigm Shift: From Tool to Collaborator}
	
	\begin{figure}[htbp]
		\centering
		\includegraphics[width=0.95\textwidth]{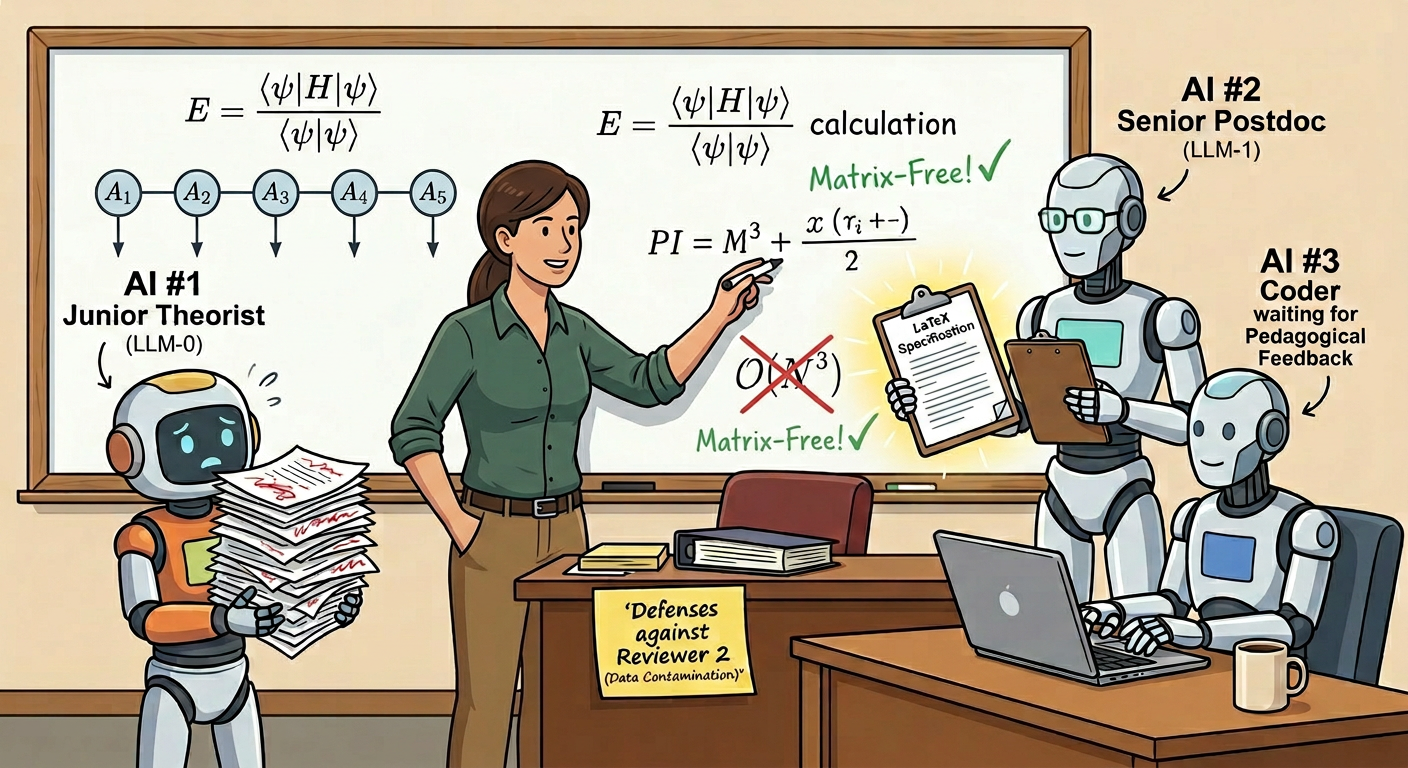} 
		\caption{\textbf{The Virtual Research Group.} To successfully write scalable quantum physics code and draft an academic manuscript, LLMs cannot be treated as magical oracles. They must be managed as a cohort of virtual students—a junior theorist for extraction, a senior postdoc for rigorous LaTeX specification, and a coder for implementation—all actively mentored and corrected by a Human Principal Investigator.}
		\label{fig:cartoon}
	\end{figure}
	
	For years, computational physicists have used computers strictly as tools—compilers, equation solvers, and numerical libraries that execute precise, deterministic commands. But over the course of 24 hours, while building a complex tensor network engine from scratch, I experienced a fundamental paradigm shift. 
	
	I had just finished a highly successful experiment using a ``Virtual Research Group'' of Large Language Models (LLMs) to translate a dense theoretical physics review into a flawless, scalable Python codebase. To achieve this, I assigned the models specific academic roles: \textbf{LLM-0} acted as the ``Junior Theorist'' extracting raw equations, \textbf{LLM-1} as the ``Senior Postdoc'' writing a mathematically rigorous LaTeX blueprint, and \textbf{LLM-2} as the ``Coder'' executing the final Python implementation. The results of this multi-agent workflow were staggering: completing a task that traditionally takes graduate students months, in under a day. 
	
	But when it came time to write the manuscript detailing this breakthrough, I realized I couldn't just open a blank LaTeX file and start typing. Nor could I simply open ChatGPT and command: \textit{``Write a 5-page physics paper about my coding experiment.''} If I did that, the AI would generate a generic, overly enthusiastic, and scientifically shallow draft—the academic equivalent of a hallucination. 
	
	To produce a manuscript worthy of \textit{Physical Review}, I had to treat the AI writing assistant not as a magical text-generator, but as a highly capable, yet inexperienced, Junior Collaborator. I had to provide the structure, enforce the scientific rigor, and define the precise narrative constraints. What follows is a behind-the-scenes look at how I co-authored a physics paper about AI, using AI, and what this implies for the future of scientific responsibility.
	
	\section{The ``Inside-Out'' Writing Strategy}
	
	When collaborating with an AI on a manuscript, the biggest mistake a researcher can make is starting with the Introduction. If you ask an LLM to write an introduction before the core arguments are locked in, it will lose context, hallucinate the narrative arc, and rush the ending. 
	
	The very first and most critical step of this collaboration is \textit{Context Loading}. Before generating a single paragraph of prose, it was my responsibility as the human Principal Investigator to ensure the AI understood the entire scope of the research. I loaded its context window with as much specific, high-fidelity information as possible: the original theoretical idea of the paper, the step-by-step project details of the coding experiment, the mathematical LaTeX specifications, and crucially, the raw, unedited conversation transcripts from the entire development path. This meant feeding the writing AI the complete history of interactions with LLM-0 (the theory extraction), LLM-1 (the expert blueprinting), and LLM-2 (the code implementation and physics debugging). I had to explicitly ground the AI in the reality of the physics project so that it would not revert to generic, parametric generalizations.
	
	\subsection{The Conversational Genesis of Core Concepts}
	A defining characteristic of this collaboration was the conversational genesis of the paper's core vocabulary. The key terms that anchored the manuscript were not generated zero-shot by the AI, nor were they fully pre-conceived by me; instead, they emerged through iterative dialogue. 
	
	For example, after reflecting on the workflow, I prompted the AI with a raw, intuitive observation: 
	\begin{quote}
		\textbf{User:} ``In the whole work, I feel like training AI students.''
	\end{quote}
	The AI served as an intellectual sounding board, synthesizing this sentiment into the formal ``Virtual Research Group'' metaphor and explicitly assigning the academic roles of Junior Theorist, Senior Postdoc, and Coder to the respective LLM stages. Similarly, when I emphasized to the AI that the intermediate LaTeX note was the decisive factor in achieving cross-model reproducibility, the AI proposed conceptualizing the LaTeX specification as a ``Universal API.'' This dynamic illustrates the true power of LLMs in academic writing: they can elevate a researcher's raw physical intuition into polished, high-impact scientific paradigms through conversational symbiosis.
	
	\subsection{Drafting the Blueprint}
	Only after these logical pillars were rigidly defined through discussion did we iterate on the title, eventually landing on: \textit{``From Paper to Program: A Multi-Stage LLM-Assisted Workflow for Accelerating Quantum Many-Body Algorithm Development.''} With the blueprint secured and the context fully loaded, we began drafting the manuscript section by section. But as the drafting commenced, it became immediately clear that the AI, left to its own devices, lacked the nuanced physical intuition and academic diplomacy required for a top-tier journal. I had to step in.
	
	\section{Mentoring the AI: Enforcing Academic Rigor}
	
	The true value of Human-in-the-Loop (HITL) methodology is not fixing typos; it is enforcing domain-specific scientific rigor and logical consistency. Here are three exact moments from our chat transcripts where human responsibility was required to save the manuscript.
	
	\subsection{Catching Physics Inaccuracies (Discrete vs. Continuous)}
	While drafting the Introduction, the AI attempted to contrast abstract math with actual code. It generated the following sentence: \textit{``However, translating the continuous, analytical mathematics of tensor network theory into discrete, high-performance software remains a formidable challenge.''}
	
	As a physicist, I immediately flagged this logical flaw. Standard tensor networks operate on discrete spin lattices, not continuous space. I prompted the AI:
	\begin{quote}
		\textbf{User:} ``In the introduction... What does the continuous mean?''\footnote{To maintain complete transparency and authenticity, all user prompts quoted in this essay are reproduced verbatim from the original raw transcripts, retaining all original typographical and grammatical errors. This illustrates the fast, conversational nature of human input versus the polished output of the AI.}
	\end{quote}
	Guided by this feedback, the AI realized its categorical error and revised the text to contrast the \textit{``abstract, diagrammatic''} mathematics of tensor networks with explicit array operations—a much more precise and physically accurate framing.
	
	\subsection{Modern Condensed Matter Taxonomy}
	Later, while describing a specific quantum model (the Spin-1 AKLT chain), the AI stated that the code successfully captured the model's ``hidden topological order.'' While historically understandable, this ignores the modern classification of quantum phases. I intervened to enforce current scientific standards:
	\begin{quote}
		\textbf{User:} ``On the physical the Haldane phase is charcterized by symmetry protected topological order insteaf of topological order. So how to make modification to 'hidden topological order in the Haldane phase'?''
	\end{quote}
	The AI updated the manuscript, replacing all generic topological references with the strict, modern taxonomy: \textit{Symmetry-Protected Topological (SPT) order}.
	
	\subsection{Academic Diplomacy and Professional Tone}
	In the Discussion section, the AI correctly identified that zero-shot coding fails because an LLM's pre-training data is a messy mix of different open-source conventions. However, it drafted a highly aggressive sentence: \textit{``...the vast, uncurated amalgamation of diverse tensor network libraries (e.g., ITensor, TeNPy) present in their pre-training data.''}
	
	Knowing that the creators of these libraries are highly respected pillars of the computational physics community, I exercised my responsibility as the human author to manage the paper's academic diplomacy:
	\begin{quote}
		\textbf{User:} ``Will this statemement, "In zero-shot physics coding, LLMs rely heavily on their parametric memory—the vast, uncurated amalgamation of diverse tensor network libraries... upset the authors of ITensor and TeNPy?''
	\end{quote}
	The AI recognized the diplomatic faux pas. We revised the paragraph to flatter the \textit{``superb, highly optimized open-source frameworks,''} while correctly shifting the blame entirely onto the LLM's own ``convention mixing'' and hallucinatory retrieval. 
	
	\section[Anticipating ``Reviewer 2'': Closing Logical Loopholes]{Anticipating ``Reviewer 2'': Closing Logical Loopholes\footnote{In academic publishing lore, ``Reviewer 2'' is the archetypal peer reviewer known for being exceptionally critical, skeptical, and demanding of rigorous proof to close any perceived methodological loopholes.}}
	
	A crucial part of human co-authorship is anticipating skepticism. A seasoned reviewer in any high-impact computational journal will not simply accept that an AI wrote a complex codebase; they will actively look for logical flaws, data contamination, and imprecise terminology. It was my responsibility to ensure these defenses were woven directly into the manuscript.
	
	\subsection{The ``Data Contamination'' Defense}
	The most obvious critique of AI-generated code is the possibility that the model is simply regurgitating memorized, open-source scripts (e.g., from GitHub repositories like TeNPy or ITensor) rather than dynamically reasoning from the provided LaTeX specification. 
	
	I directed the AI to explicitly close this loophole in the text. We highlighted that the generated Python code utilized highly idiosyncratic array manipulation strings exactly as they were newly defined in our intermediate LaTeX blueprint. Furthermore, we noted that while all models regurgitated standard textbook equations, they diverged significantly on complex, non-standard derivations—autonomously deriving unique matrix representations. This provided empirical proof of \textit{in-context symbolic reasoning} rather than parametric memorization.
	
	\subsection{The ``Model Capability'' Paradox}
	A careful reviewer reading the methodology would inevitably spot an apparent contradiction: \textit{``If the Kimi 2.5 model struggled to account for computational realities in Stage 1 (acting as LLM-0), how could the Kimi Agent perform flawlessly when deployed as the implementation coder in Stage 3?''}
	
	I realized this was not a contradiction, but rather the ultimate proof of the paper's thesis. I prompted the AI to address this paradox directly in the Discussion section. This stark contrast isolates the true bottleneck in AI-assisted scientific programming: the failure of zero-shot coding is not due to a lack of reasoning capacity within the foundation models, but rather the absence of a constrained, step-by-step mathematical context. When provided with the formal LaTeX blueprint, the exact same model transitions from producing hallucinatory pseudo-code to generating rigorous software.
	
	\subsection{The Pedantic Reviewer (Algorithmic Rigor)}
	Reviewers in theoretical physics and computer science are notoriously meticulous about algorithmic formalism. While drafting the abstract, the AI wrote that the matrix-free solver \textit{``bypasses the prohibitive $\mathcal{O}(D^4)$ memory scaling of explicit matrix construction.''} 
	
	While colloquially understood, a strict computational reviewer would flag this as imprecise, because the exact memory scaling depends on the local physical dimension $d$ and whether the algorithm is updating one or two sites. I intervened to enforce absolute rigor:
	\begin{quote}
		\textbf{User:} ``Is this statement suitable or rigorous? 'The codebase successfully executed matrix-free Hamiltonian applications in both cases, completely avoiding the $\mathcal{O}(d^2 D^4)$ memory bottlenecks.'''
	\end{quote}
	Guided by this prompt, we refined the manuscript to explicitly distinguish between the $\mathcal{O}(d^2 D^4)$ bottleneck of single-site updates and the $\mathcal{O}(d^4 D^4)$ bottleneck of two-site updates. By anticipating this level of pedantry, we ensured the manuscript spoke the exact, rigorous language expected by referees in top-tier scientific journals.
	
	\section{Directing the Visuals: AI as an Art Director}
	
	The human-AI collaboration extended beyond drafting text and code into multi-modal visual storytelling. For the original computational physics manuscript, I required professional-grade figures to illustrate the multi-agent workflow and the accelerated 24-hour timeline. However, feeding generic prompts to an AI image generator (in this case, Nano Banana 2) predictably resulted in scientific nonsense. 
	
	Rather than fixing the generated images manually, I tasked my text-based AI co-author (Gemini 3.1 Pro Preview) with acting as an ``Art Director.'' I prompted the text AI to write highly constrained, step-by-step design briefs to feed back into the image generator. The text AI translated my abstract academic requirements into explicitly graphical prompt engineering, dictating layout, typography, and color-coding constraints (e.g., using red to highlight the flawed LLM-0 draft and green for the rigorous LLM-1 blueprint).
	
	This same multi-modal collaboration was used to create the editorial cartoon for this very essay (Fig.~\ref{fig:cartoon}). When the initial image generation hallucinated a chaotic, feed-forward neural network on the whiteboard alongside the gibberish equation $\mathcal{O}(N^3)$, the human PI had to intervene to enforce physical reality:
	\begin{quote}
		\textbf{User:} ``Do you think that this looks good? I think the diagram in the whiteboard is wrong, not for 1D system/MPS/DMRG.''
	\end{quote}
	The text AI instantly recognized the physical inaccuracy and wrote a new, strict prompt for Nano Banana:
	\begin{quote}
		\textbf{AI-Generated Design Brief:} \textit{``Crucial Details (Do Not Hallucinate Math): The whiteboard must look like real 1D quantum physics. Draw a neat, horizontal row of 5 circles connected by a single horizontal line passing through their centers... This is a 1D Matrix Product State diagram. Do NOT draw a chaotic web or neural network. Write $\mathcal{O}(D^6)$ with a large, aggressive RED 'X' crossed through it.''}
	\end{quote}
	
	By mediating the interaction between the human physicist and the image generator, the text-based LLM successfully translated abstract quantum mechanical constraints into geometric instructions. The result was a suite of visually engaging, physically accurate figures and cartoons that perfectly anchored both the physics manuscript and this reflective essay.
	
	\section{Conclusion: The Future of Authorship and Transparency}
	
	Writing a paper with AI is not about automation; it is about augmentation and iteration. I did not use an AI to write my paper for me. I collaborated with an AI to structure my thoughts, refine my logical arguments, and typeset my results. Throughout the process, the human physicist remained the Principal Investigator—setting the curriculum, correcting the physics, and ensuring the scientific truth. The human contribution has shifted from typing boilerplate text to high-level intellectual steering.
	
	As this ``Virtual Research Group'' paradigm becomes the standard for scientific software development and manuscript drafting, it introduces profound questions regarding academic integrity. If LLMs are actively contributing to the structural and syntactical generation of scientific literature, how do we evaluate the origin of the ideas? How do we hold authors accountable?
	
	The answer lies in radical transparency. It is no longer sufficient to simply state in an acknowledgment section, \textit{``ChatGPT was used to improve readability.''} We must treat interactions with AI agents as raw experimental data. Therefore, I propose that \textbf{authors must be required to include full, unedited transcripts of their AI interactions as supplementary material} if AI tools were utilized in a manuscript's preparation. 
	
	By publishing our prompts and the AI's iterative responses, we demystify the ``black box'' of generative AI. We prove that the human researcher was actively steering the logic, catching the errors, and driving the innovation. In the age of AI, transcript transparency is the only way to preserve the accountability of authorship and the integrity of the scientific record.
	
	\section*{Acknowledgments}
	
	I would like to explicitly acknowledge my AI collaborator—specifically Gemini 3.1 Pro Preview—for acting as a tireless virtual postdoc and editorial assistant throughout this writing project. This essay, much like the physics manuscript it describes, was co-authored through an iterative, conversational process. The AI provided the structural organization, syntax generation, and rapid iteration, while I provided the physical intuition, academic direction, and pedagogical feedback. This work stands as a testament to the profound potential of human-AI symbiosis in advancing scientific research. I also formally acknowledge the foundation models (Kimi 2.5, Gemini 3.1 Pro Preview, GPT 5.4, and Claude Opus 4.6) that constituted the ``Virtual Research Group'' making the underlying accelerated coding workflow possible.
	
	\section*{Data and Code Availability}
	
	Practicing the radical transparency advocated in this essay, all materials associated with the drafting of both the physics manuscript and this reflective essay have been made publicly available. To ensure full reproducibility and academic accountability, the complete, unedited transcripts of the human-AI interactions are hosted in the GitHub repository: \url{https://github.com/yizhou76-sudo/Essay-Write-with-AI}. Readers are encouraged to review the prompt history to independently verify the human-in-the-loop steering, logical corrections, and academic mentorship that shaped the final publications.
	
\end{document}